\documentclass[prd, aps]{revtex4}

\usepackage{latexsym, graphicx}
\begin{document}
\title{Radiation by an Unruh-DeWitt Detector in Oscillatory Motion\footnote{Prepared for {\it Proceedings of the 14th 
Marcel Grossman Meeting on General Relativity, July 12-18, 2015}, edited by Massimo Bianchi, Robert T Jantzen, Remo Ruffini
(World Scientific, Singapore, 2016).}}
\author{Shih-Yuin Lin$^*$}
\email{sylin@cc.ncue.edu.tw}
\affiliation{Department of Physics, National Changhua University of Education,
Changhua 50007, Taiwan}
\affiliation{Department of Physics and Astronomy, University of Waterloo,
Waterloo, Ontario N2L 3G1, Canada}
\affiliation{Perimeter Institute for Theoretical Physics,
Waterloo, Ontario N2L 2Y5, Canada}
\begin{abstract}
Quantum radiated energy flux emitted by an Unruh-DeWitt (UD) detector, with the internal harmonic oscillator coupled to a massless scalar field, in linear oscillatory motion in (3+1) dimensional Minkowski space is studied by numerical methods. Our results show that quantum interference can indeed suppress the signal of the Unruh effect if the averaged proper acceleration is sufficiently low, but not in the regime with high averaged acceleration and short oscillatory cycle. While the averaged radiated energy flux over a cycle is always positive as guaranteed by the quantum inequalities, an observer at a fixed angle may see short periods of negative radiated energy flux in each cycle of motion, which indicates that the radiation is squeezed. This reveals another resemblance between the detector theory and the moving-mirror model.
\end{abstract}
\date{January 21, 2016}

\maketitle

\section{Introduction}

The Unruh effect states that a uniformly, linearly accelerated detector in the Minkowski vacuum will experience thermal fluctuations at a temperature proportional to the proper acceleration of the detector~\cite{Un76}, called the Unruh temperature. While the derivation is well established, direct observation of the Unruh effect in laboratories is still lacking. One possible experimental evidence is the 
electron depolarization in storage rings, namely, the Sokorov-Ternov effect~\cite{ST63}, which can be directly connected to the ``circular 
Unruh effect"~\cite{BL83}. Nevertheless, the centripetal acceleration in the circular Unruh effect is quite different in nature from the 
original linear, uniform acceleration in the Unruh effect~\cite{HR00}. To get closer to the original conditions in Unruh's derivation,  
people have proposed to look at the correction by the Unruh effect to the radiation emitted by {\it linearly} 
accelerated charges or atoms~\cite{CT99, SKB03, SSH06}, which is called the ``Unruh radiation".

Seeking the imprint of the Unruh temperature in quantum radiation is not as straightforward as it appears.
To define a finite temperature in a detector-field state, one needs a detector in equilibrium with the field. Unfortunately, a uniformly, linearly accelerated detector (analogous to an atom) coupled to a massless scalar field in (1+1)D Minkowski space emits no radiation in {\it equilibrium} conditions~\cite{Gr86, HR00}. In (3+1)D Minkowski space there will be radiations by an Unruh-DeWitt (UD) 
detector~\cite{Un76, DeW79} in steady state at late times (with a constant radiating rate with respect to the proper time of the 
detector), but the radiated energy is not converted from the energy flux experienced by the detector in the Unruh effect~\cite{LH06}. 
The interference between the vacuum fluctuations driving the detector and the radiation emitted by the driven detector is perfectly destructive in steady state. 

In laboratories, however, it is impossible to produce any eternal constant linear acceleration for a charge or an atom. For example, in 
Ref.~\cite{CT99} the authors proposed to use an intense laser field to drive the charge motion, meaning that the acceleration is 
linear, oscillatory, but not uniform. %
A positive aspect of similar laser-driven experiments is that the charges or atoms 
are never in equilibrium conditions, so the quantum radiation may not be totally suppressed by interference.
Moreover, while temperature is not well defined in this non-equilibrium setup, we can still define a time-varying effective temperature 
whose value is close to the Unruh temperature with the averaged acceleration of the oscillatory motion~\cite{DLMH13}. 
Indeed, as we will demonstrate later, 
in the regime of high acceleration and short oscillating cycle of the motion, 
the signal of the effective Unruh temperature can be pronounced in the quantum radiation. 
Further, when observed in a fixed angle, there will be negative radiated energy flux in short periods during each cycle of oscillatory motion. This indicates that the Unruh radiation observed at the null infinity corresponds to a 
squeezed state of the field~\cite{UW84, SSH06}. 

We show our numerical results in Section 4, before which some technical issues are addressed in Sections 2 and 3.
Our model is the UD detector theory considered in Refs.~\cite{LH06} and~\cite{DLMH13}: the {\it internal} harmonic 
oscillator is minimally coupled to a massless scalar field at the position of the point-like detector, 
while the worldline of the detector is not determined dynamically but given by a classical 
solution in electrodynamics with the radiation reaction neglected. 
We take $c=\hbar=1$ and signature $(-,+,+,+)$.

\section{Renormalized Expectation Values of Stress-Energy Tensor}

As we showed in Ref.~\cite{LH06}, by virtue of the linearity of the UD detector theory, the field operator $\hat{\Phi}^{}_x$ after the 
detector-field coupling is switched on will become a linear combination of the mode functions each associated with a creation 
($\hat{b}^\dagger_k$) or annihilation operator ($\hat{b}^{}_k$) of the free field, or a raising ($\hat{a}^\dagger_A$) or lowering 
operator ($\hat{a}^\dagger_A$) of the free detector $A$. Also due to the linearity, a mode function of the field has the form 
$\phi^\alpha_x = \phi^{{}^{[0]}\alpha}_{\; x} + \phi^{{}^{[1]}\alpha}_{\;x}$ ($\alpha = A, k$; $\phi^A_x$ and $\phi^k_x$ are associated 
with $\hat{a}^{}_A$ and $\hat{b}^{}_k$, respectively), which is the superposition of the homogeneous solution $\phi^{{}^{[0]}\alpha}_x$ 
and the inhomogeneous solution $\phi^{{}^{[1]}\alpha}_{\;x}$ sourced from the point-like detector. One can group the homogeneous solutions 
of the mode functions with the associated operators into $\hat{\Phi}^{{}^{[0]}}_x$ and the inhomogeneous solutions into 
$\hat{\Phi}^{{}^{[1]}}_x$ such that the field operator is in the form $\hat{\Phi}^{}_x=\hat{\Phi}^{{}^{[0]}}_x+\hat{\Phi}^{{}^{[1]}}_x$. 
Suppose right before the initial moment $t=0$ when the detector-field coupling is switched on, the combined system is in the factorized 
state $|\psi(0)\rangle = |g^{}_A\rangle \otimes | 0^{}_M\rangle$,  
which is a product of the ground state of the free UD detector $|g^{}_A\rangle$ and the Minkowski vacuum of the field $|0^{}_M\rangle$.
Then the correlators of the field amplitude at different spacetime points $x$ and $x'$ are given by
\begin{equation}
   G(x,x')\equiv \langle \psi(0)| \hat{\Phi}^{}_x \hat{\Phi}^{}_{x'} |\psi(0)\rangle = \sum_{i,j=0,1} G^{(ij)}(x,x'),
\end{equation}
where 
$G^{(ij)}(x,x') \equiv \langle \hat{\Phi}^{{}^{[i]}}_x \hat{\Phi}^{{}^{[j]}}_{x'} \rangle$
with respect to the initial state $|\psi(0)\rangle$. 
The expectation value of the stress-energy tensor (minimal, $\xi=0$) can be written as
\begin{equation}
  \langle T_{\mu\nu}[\Phi(x)]\rangle = 
	\lim_{x'\to x} \left[ {\partial\over\partial x^\mu}{\partial\over\partial x'^\mu}
	- {1\over 2} g^{}_{\mu\nu}g_{}^{\rho\sigma}{\partial\over\partial x^\rho}{\partial\over\partial x'^\sigma}\right]G(x,x')
	\equiv \langle T^{(ij)}_{\mu\nu}(x)\rangle
\end{equation}
where $\langle T^{(ij)}_{\mu\nu}\rangle$ is contributed by $G^{(ij)}$.
$G^{(00)}(x, x')$ is the Green's function of the free field, diverges as $x'\to x$, so does $T^{(00)}_{\mu\nu}$.
Nevertheless, there is no physical effect from this part of the stress-energy tensor in Minkowski space, and so it can be subtracted by 
introducing the normal ordering of the creation/annihilation or raising/lowering operators. We thus define the renormalized stress-energy 
tensor as
$\langle T_{\mu\nu}(x) \rangle_{ren} \equiv
\langle :T_{\mu\nu}(x): \rangle = \langle T_{\mu\nu}(x) \rangle -\langle T_{\mu\nu}^{(00)}(x)\rangle$.
Doing this is nothing but setting the zero point of vacuum energy. 

Suppose a UD detector is oscillating about the origin in space.
Suppose a set of the radiation-detecting apparatus are sitting at large constant radius $r$ at different angles from the spatial origin
of the laboratory frame, namely, located at $x^\mu$ in the Minkowski coordinates with $(x^1, x^2, x^3)= (r \sin\theta \cos \varphi, \,
r \sin\theta \sin \varphi, \, r \cos \theta)$. 
Then the angular distribution of the radiated energy flux measured in laboratory can be divided into two parts,
\begin{equation}
  {d{\cal P} \,\over d\Omega^{}_{\rm II}}(t_0,\theta,\varphi) = -\lim_{r\to\infty} r^2 \langle T_{tr}(x) \rangle_{ren}   
	= {d{\cal P}^{\{01\}}\over d\Omega_{\rm II}} + {d{\cal P}^{(11)}\over d\Omega_{\rm II}}  \label{dPdWIIren}
\end{equation}
at the coordinate-time $x^0 = t_0 + r$ of the apparatus, where $t_0$ is the time when the radiation is emitted by the UD detector
in view of the Minkowski observer.
Here
\begin{equation}
	{d{\cal P}^{(01)}\over d\Omega_{\rm II}} \equiv 
	-\lim_{r\to\infty}{r^2\over 2}\left\{ \left(\partial_t \partial_{r'} + \partial_r \partial_{t'}\right)
	 \left[G^{(10)}
	              (x,x')+ G^{(01)}
								(x,x')\right]\right\}
\end{equation}
is contributed by the interference between vacuum fluctuations and the backreaction of the detector driven by vacuum fluctuations, and
\begin{equation}
	{d{\cal P}^{(11)}\over d\Omega_{\rm II}} \equiv -\lim_{r\to\infty,x'\to x}{r^2\over 2}  
	\left(\partial_t \partial_{r'} + \partial_r \partial_{t'}\right) G^{(11)}(x,x') \label{dP11dW}
\end{equation}
is contributed purely by the retarded solution emitted by the detector. 
From (A1) in Ref.~\cite{LH06}, one has 
\begin{eqnarray}
	\partial_t \partial_{r'} G^{(11)}(x,x')
	&=& {\lambda_0^2\over (2\pi)^2 4 {\cal R R'}}\theta(\eta_-)\theta(\eta_-') \times \nonumber\\ & &
  \left[ {{\cal R}_{,t}{\cal R}'_{,r'}\over {\cal R R'}}\langle Q(\eta_-)Q(\eta_-') \rangle +
  \eta_{-,t}\eta'_{-,r'}\langle P(\eta_-)P(\eta_-')\rangle\right. \nonumber\\ & & 
	\left. - {{\cal R}_{,t}\over {\cal R}}\eta'_{-,r'} \langle Q(\eta_-)P(\eta_-')\rangle -
	\eta_{-,t}{{\cal R}'_{,\nu}\over {\cal R'}}\langle P(\eta_-)Q(\eta_-')\rangle \right]. \label{DDG11}
\end{eqnarray}
with the retarded distance $aX/2$ and the retarded proper time $\eta_- = \tau_- -\tau_0$ for a uniformly accelerated detector there 
being generalized to ${\cal R}$ and $\eta_-$ for a detector in oscillatory motion\footnote{In Eq. (A1) in Ref.~\cite{LH06}, 
$\partial_t \partial_{r'} G^{(11)}_{\rm v}(x,x')$ only counts the contribution by the 
v-parts of the correlators of the detector $\langle .. \rangle_{\rm v}$. The expression for the a-part, $\partial_t \partial_{r'} 
G^{(11)}_{\rm a}(x,x')$, has the same form as Eq. (A1) in Ref.~\cite{LH06} except the v-parts of the detector-detector correlators
$\langle .. \rangle_{\rm v}$ are replaced by the a-parts, $\langle .. \rangle_{\rm a}$. Since $\langle \Phi_x \rangle =0$ in the cases
we are considering, the expression for $\partial_t \partial_{r'} G^{(11)}(x,x') =\partial_t \partial_{r'} [G^{(11)}_{\rm a}(x,x') + 
G^{(11)}_{\rm v}(x,x') ]$  is simply the same expression as Eq. (A1) in Ref.~\cite{LH06} except all the v-part of the detector-detector
correlators there are replaced by the complete one, namely, 
$\langle .. \rangle_{\rm v} \to \langle .. \rangle_{\rm v}=\langle .. \rangle_{\rm a}+ \langle .. \rangle_{\rm v}$.}.
${\cal R}$ is the retarded distance determined by the local frame of the detector. 
For an observer at the null infinity, the more the 4-velocity of the detector is pointing towards the observer,  the smaller 
${\cal R}/r$ is. 

Below, we are considering the cases with the detector in oscillatory motion in the $x^3$-direction, namely, $z^\mu(\tau) = (z^0(\tau), 
0,0, z^3(\tau))$. The radiation will be independent of the azimuth angle $\varphi$, and so we are looking at the time evolution of
the polar-angular distribution of the radiated energy flux
\begin{equation}
  {d{\cal P}\over d\theta} =  \int_0^{2\pi} d\varphi {d{\cal P}\,\over d\Omega_{\rm II}},
\end{equation}
observed by the apparatus. The total radiated energy flux is thus
 ${\cal P}=  \int_0^{\pi} \sin\theta d\theta {d{\cal P}\over d\theta}$.

\section{Subtraction of Singularities}

One could extract the Unruh temperature from the correlators of the detector in (\ref{DDG11}) and thus (\ref{dP11dW}), if measurable. 
So we call the all-retarded-field part of the radiated energy flux $d{\cal P}^{(11)}/d\Omega_{\rm II}$ as the naive Unruh radiation.
It diverges when one takes the coincidence limit on the two-point correlators of the detector ($\lim_{\tau'_-\to \tau_-} \langle 
\hat{Q}(\tau_-),\hat{Q}(\tau'_-)\rangle$, $\langle \hat{Q}(\tau_-),\hat{P}(\tau'_-)\rangle$, and $\langle \hat{P}(\tau_-),\hat{P}(\tau'_-)
\rangle$). This is originated from the singular behavior of the Wightman function of the field in the double integral in calculating  
the two-point correlators. When the trajectory of the detector is not as simple as those in uniform motion or uniform acceleration, 
setting cutoffs in the double integral for the correlators is not trivial. In Refs.~\cite{DLMH13} and \cite{OLMH12} we 
have dealt with these singularities carefully. We subtract the integral for the two-point correlators of the detector in oscillatory motion by those for a uniformly accelerated detector. The subtracted integral gives a finite result. Then we add the analytic results for the uniformly accelerated detector back, whose singular behavior are well understood and under control once the UV cutoff is introduced.

For the interference part of the radiated energy flux, $d{\cal P}^{(01)}/d\Omega_{\rm II}$, the situation is similar.
Substitute the general solutions for the mode functions into the interference term of the field-field correlators, one obtains
\begin{eqnarray}
	& & G^{(01)}_{\rm v}(x,x') + G^{(10)}_{\rm v}(x,x') \nonumber\\ &=& 2 {\rm Re}
	  {2\hbar\gamma\theta(\eta_-(x))\over 4 \pi \Omega {\cal R}(x)} \int_{\tau_0}^{\tau_-(x)} d\tilde{\tau} 
		K(\tau_-(x) -\tilde{\tau}) D^+(x', z(\tilde{\tau}-i\epsilon)) 
\end{eqnarray}
where $K(x)\equiv e^{-\gamma x}\sin\Omega x$ and the Wightman function of our free massless scalar field is 
   $ D^+(x, x') = \hbar[(2\pi)^2 (x_\mu - x'_\mu)(x^\mu - x'^\mu)]^{-1}$
with a proper choice of the integration contour understood.
In calculating the interference terms for $-r^2 \langle T_{tr}\rangle$, the singularity arises in the terms containing
\begin{eqnarray} 
  {\partial\over\partial x^\nu} D^+(x - z(\tau-i\epsilon)) = {\hbar\over 2\pi^2} {z_\nu(\tau-i\epsilon)-x_\nu \over
	  \left[(x_\mu - z_\mu(\tau-i\epsilon))(x^\mu - z^\mu(\tau-i\epsilon))\right]^2},
\end{eqnarray}
whose denominator vanishes when $\tau=\tau_-(x)$ and $\epsilon\to 0+$ (actually we use this condition to determine the retarded
time $\tau_-(x)$). When $\epsilon$ is positive and non-zero, expanding $z^\mu(\tau_- -i\epsilon) \approx z^\mu(\tau_-)-i\epsilon 
\dot{z}^\mu(\tau_-) + (-i\epsilon)^2 \ddot{z}^\mu(\tau_-)/2 + \cdots$, one has
\begin{eqnarray}
  & &\partial_\nu D^+(x - z(\tau_-(x) -i\epsilon)) = {\hbar\over 2\pi^2}\left\{ 
	    {1\over\epsilon^2} {x_\nu-z_\nu^-\over 4\left[ \dot{z}_\mu^-(x^\mu -z^\mu_-)\right]^2} + \cdots \right\}
\end{eqnarray}
with $z^\mu_- \equiv z^\mu(\tau_-(x))$. To subtract out the divergent $\epsilon^{-2}$ and $\epsilon^{-1}$ terms,
one needs to introduce a reference worldline  $\tilde{z}^\mu(\tau)$ with $\tilde{z}^\mu(\tau_-) = z^\mu(\tau_-)$, $\dot{\tilde{z}}_{\rm r}^\mu
(\tau_-)=\dot{z}^\mu(\tau_-)$, and $\ddot{\tilde{z}}_{\rm r}^\mu(\tau_-)=\ddot{z}^\mu(\tau_-)$. 
For a general worldline, the simplest reference worldlines for subtraction are those for uniformly accelerated detectors, and
luckily, we have obtained the analytic results of the interference term for the uniformly accelerated detector in close forms in 
Ref.~\cite{LH06}. Similar to what we did for 
$d{\cal P}^{(11)}/d\Omega_{\rm II}$, after we get the finite result for the subtracted interference term
$d{\cal P}^{(01)}/d\Omega_{\rm II}|_{z^\mu} -d{\cal P}^{(01)}/d\Omega_{\rm II}|_{\tilde{z}^\mu}$, we add the analytic result back in the final step to get the complete result with the divergences well controlled. 

Actually, for the reference worldlines in uniform acceleration, the UV divergence in $G^{(11)}$ will be exactly canceled by the ones in 
the interference terms $G^{(10)}+G^{(01)}$, as shown in Ref.~\cite{LH06}.  
Thus, combining the numerical result of the subtracted energy flux and the exact 
analytic result from the reference worldlines, the final result will be regular and independent of the UV cutoff.

\section{Numerical Results}

\begin{figure}
\includegraphics[width=5.5cm]{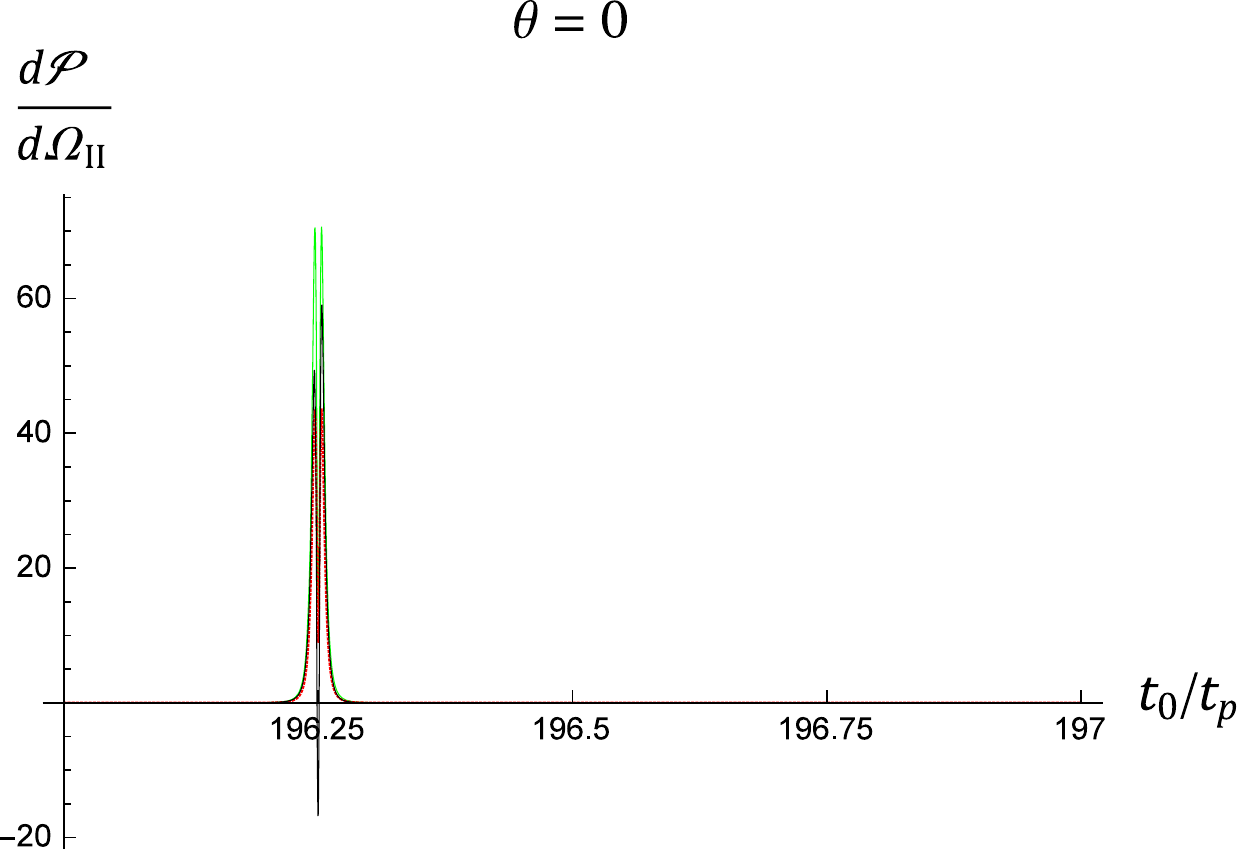}
\includegraphics[width=5.5cm]{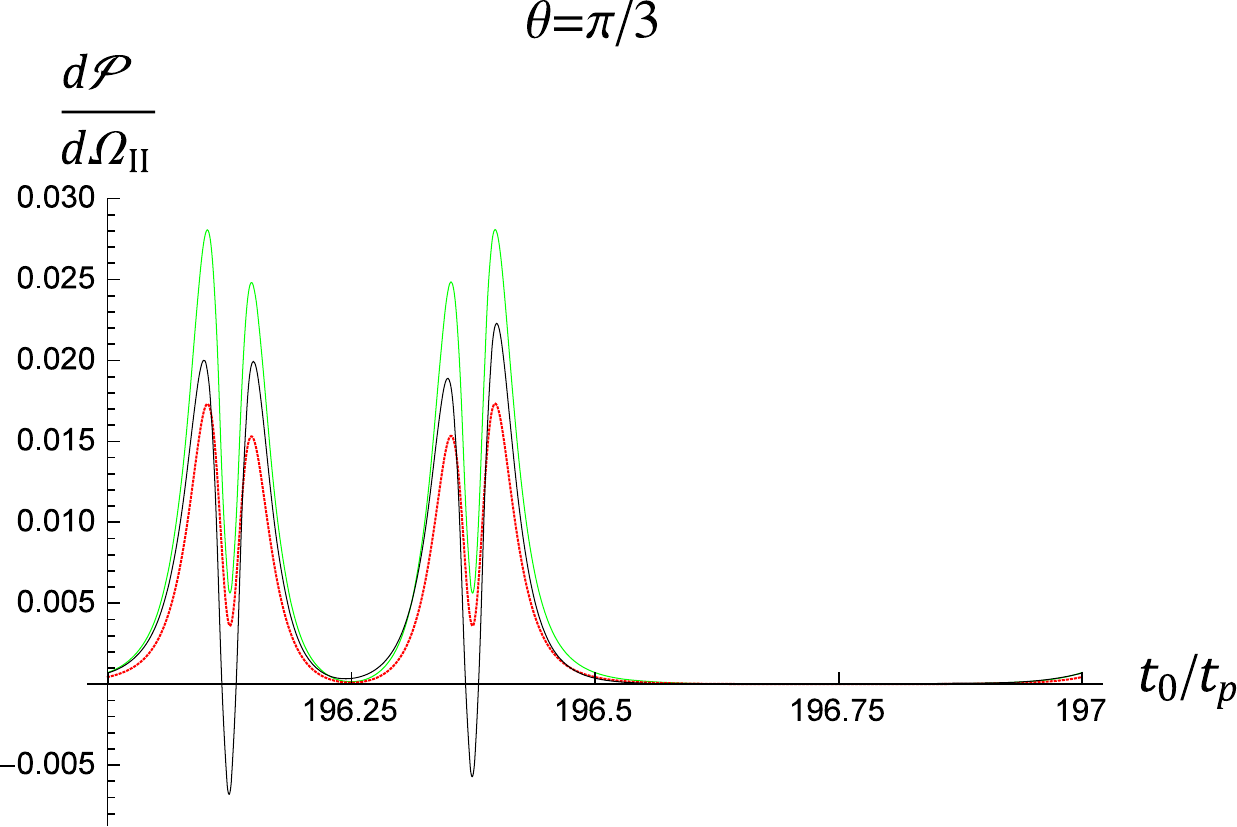}
\includegraphics[width=5.5cm]{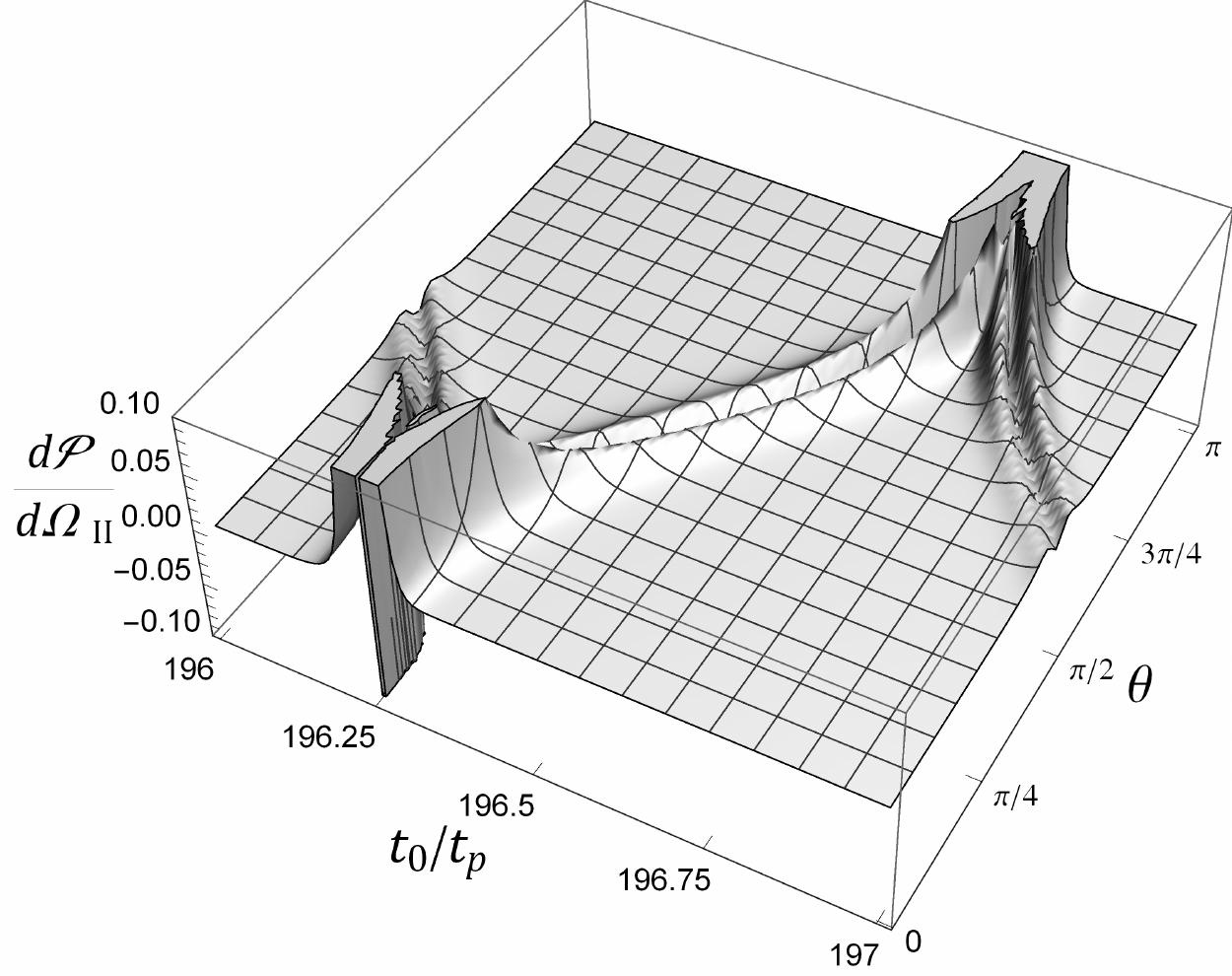}\\
\includegraphics[width=5.5cm]{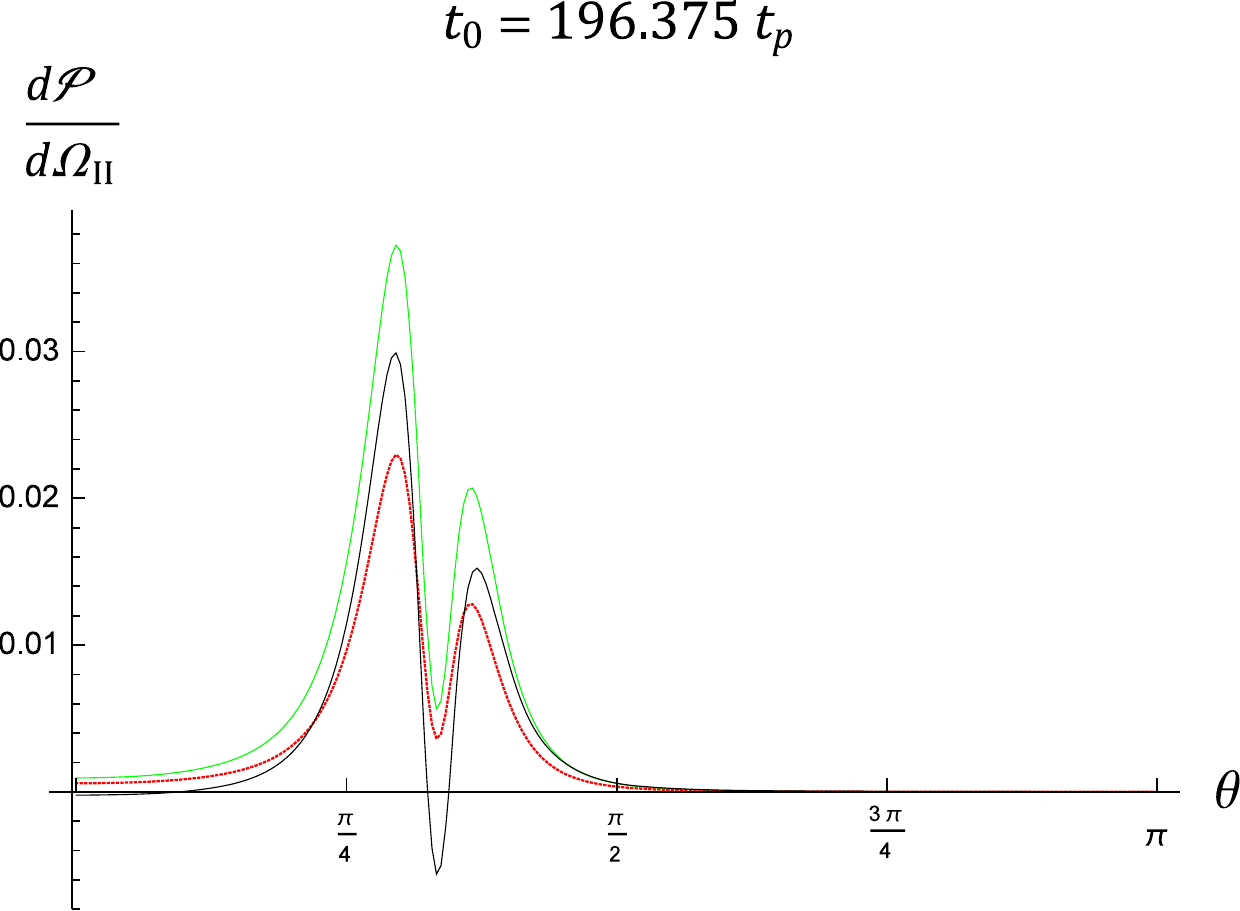}
\includegraphics[width=5.5cm]{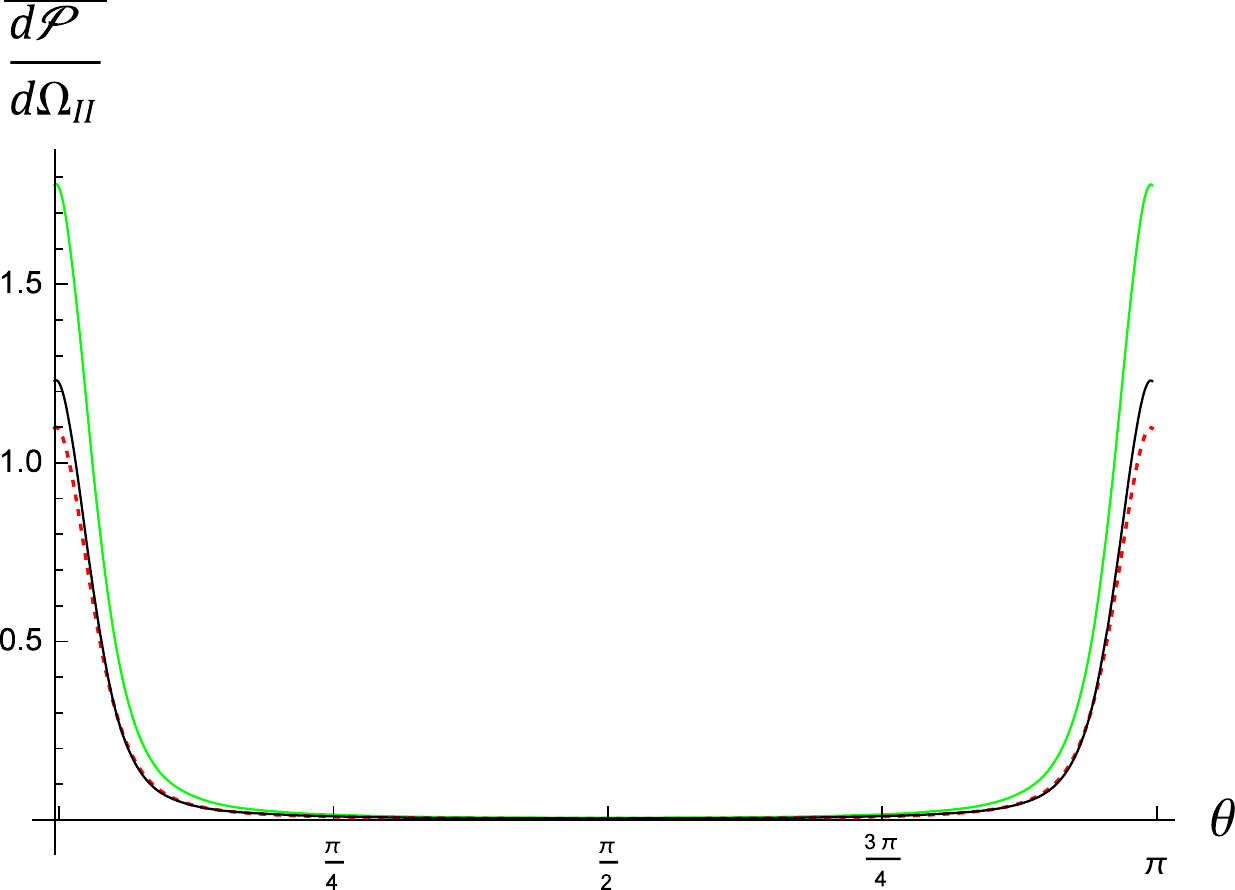}
\caption{The radiated energy flux $d{\cal P}/d\Omega_{\rm II}$ against $t_0/t_p$ at fixed $\theta$ 
(upper-left and upper-middle) and against $\theta$ at fixed $t_0$ defined below (\ref{dPdWIIren}) (lower-left), 
emitted by a detector with the coupling strength $\gamma=0.01$ and the natural frequency of the detector $\Omega=2.3$, 
along the worldline (\ref{CTtraj}) with $\omega=3.277$, $a_0=2$, and so the averaged proper acceleration $\bar{a}=10$,
the period of each cycle $t_p\equiv 2\pi/\omega=1.917$ in the rest frame, and $\tau_p = 0.838$ in the comoving frame~\cite{DLMH13}. 
The black and green solid curves represent the energy flux with and without the interference terms, respectively, while the red-dotted 
curves represent the energy flux with no interference or the Unruh effect, namely, the two-point correlators for the accelerated detector 
have been replaced by the ones for an inertial detector. 
The lower-right plot is the energy flux averaged over a cycle against the observing angle $\theta$, which is positive 
for all $\theta$. The radiated energy is concentrated around $\theta=0$ and $\pi$, the directions of the linear oscillatory motion.}
\label{CTrad}
\end{figure}

\begin{figure}
\includegraphics[width=5.5cm]{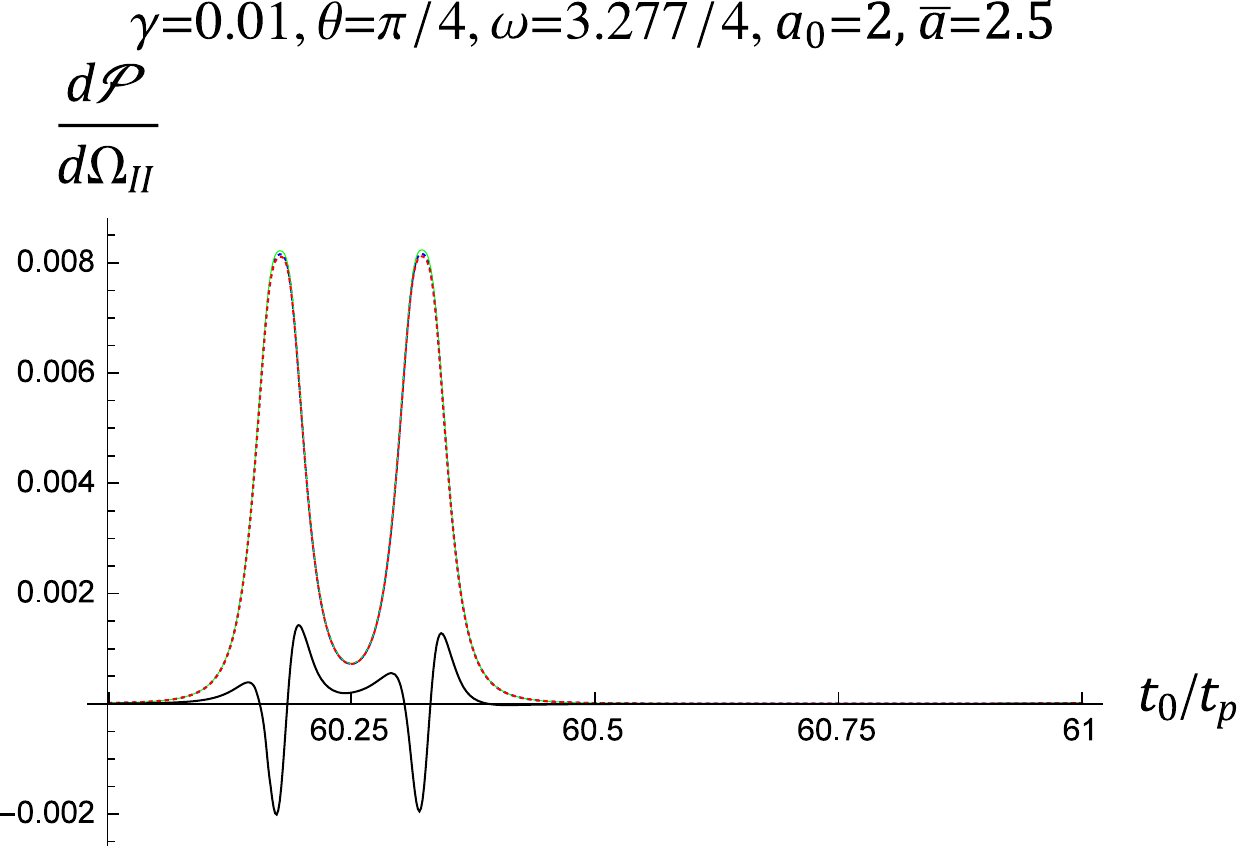}
\includegraphics[width=5.5cm]{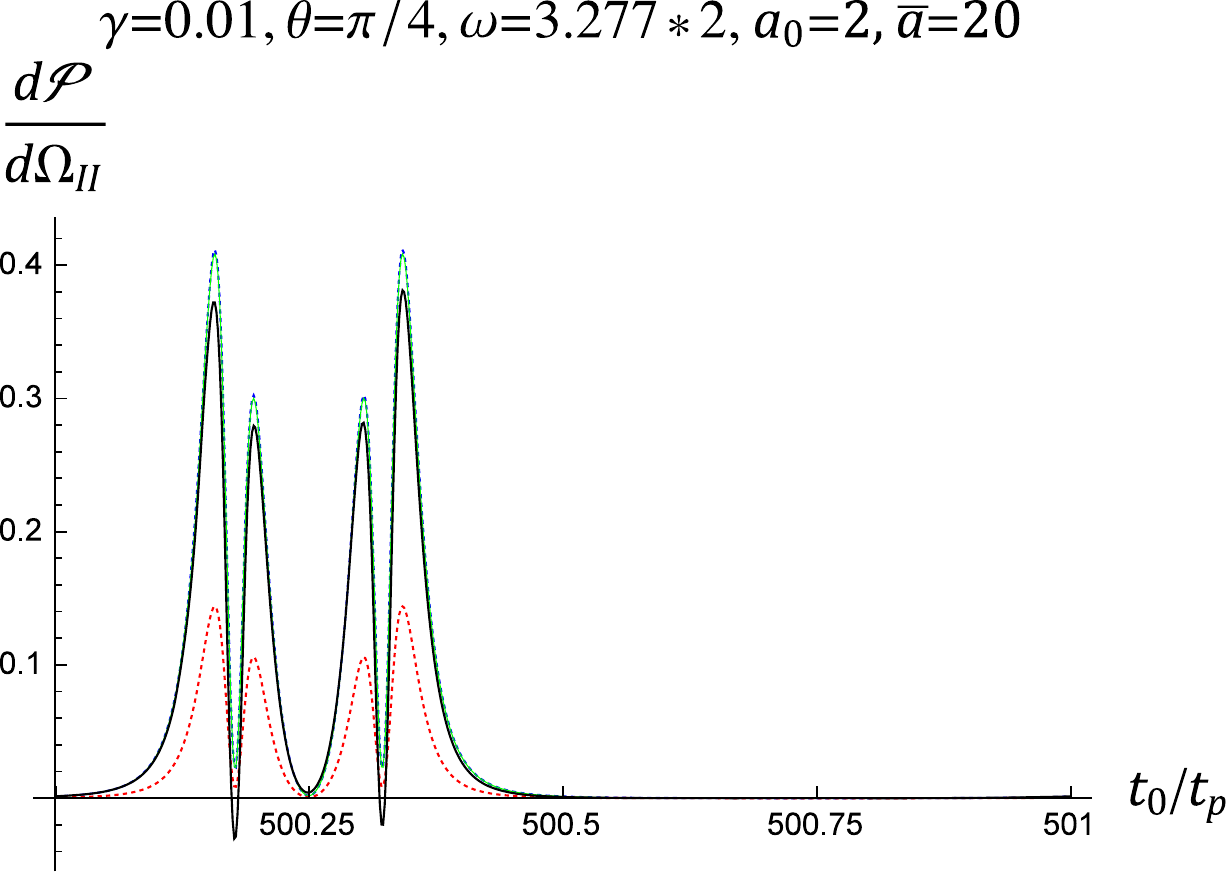}
\caption{Time evolution of the radiated energy flux at $\theta=\pi/4$ with the averaged acceleration at low ($\bar{a}=2.5$, left plot) 
and high ($\bar{a} = 20$, right) values. 
We compare the full result $d{\cal P}/d\Omega_{\rm II}$ (black) with the naive results $d{\cal P}^{(11)}/d\Omega_{\rm II}$:
the green, red dotted, and blue dotted curves are contributed by the correlators of a detector in oscillatory motion,
at rest, and in uniform acceleration with proper acceleration $\bar{a}$, respectively.}
\label{CTvarw}
\end{figure}

As an example, we consider a worldline given in Ref.~\cite{CT99},
\begin{equation}
  z^\mu_{CT}(t) = \left( t, 0, 0, -{1\over\omega}\sin^{-1}{2a_0\cos\omega t \over \sqrt{1+4 a_0^2}}\right) ,
	\label{CTtraj}
\end{equation}
which is the trajectory of a charge at a nodal point of magnetic field in a cavity. Our numerical results are shown in Figures
\ref{CTrad} and \ref{CTvarw}.

The effective temperature~\cite{DLMH13} in the example in Figure \ref{CTrad} is about $T_{\rm eff}\approx 1.6754$ to $1.6767$,
which is higher than the naive Unruh temperature $\bar{a}/(2\pi) \approx 1.5916$~\cite{DLMH13}. 
while in Figure \ref{CTvarw} (right), the effective temperature is about $3.156$, which is lower than $\bar{a}/(2\pi) \approx 3.183$.  
The deviations from the naive Unruh temperature $\bar{a}/(2\pi)$ due to non-uniform accelerations are, however, negligible in our results
of the radiated energy flux, especially during the times when the flux is negative. They are more significant when the radiated energy 
flux reaches the peak values. One may substitute an effective acceleration $a_{\rm eff} = 
2\pi T_{\rm eff}$, instead of $\bar{a}$, into the correlators for a uniformly accelerated detector to get a better estimate for the naive 
radiation.

In the upper row of Figure \ref{CTrad}, one can see that the time evolution of the full radiated energy flux at a fixed angle 
has a negative period between two positive main peaks.
At $\theta=0$, this occurs around the retarded time when the direction of the detector's acceleration is switching.
At each fixed time, the radiated energy flux always become negative around some observing angle, as shown in Figure \ref{CTrad} 
(lower-left). This negative energy flux reveals another resemblance between the detector theory and the 
moving mirror models in quantum field theory in curved spacetime: it is well known that in the moving-mirror models in (1+1)D 
a similar negative energy flux will arise if the acceleration of the mirror is non-uniform~\cite{FD76}\footnote{Actually, 
the Unruh-DeWitt-like detector theory in (1+1)D has been used to describe the physical mirrors~\cite{SLH15}, 
which generate the boundary conditions for the field dynamically at their positions.}.
The radiation with negative energy here does not imply absorption, since it can still excite a UD detector, and produce 
no decrease of entropy~\cite{DOS82}. The presence of the short negative energy flux simply indicates that the Unruh radiation corresponds 
to a multi-mode squeezed state of the field~\cite{SSH06}. 
The averaged energy flux over a cycle of oscillation at each fixed angle, or over all solid angle at each fixed time, must be positive
(Figure \ref{CTrad} (lower-right)), as guaranteed by the quantum inequalities~\cite{Fo91}.

The radiation by the detector should cease as its averaged acceleration $\bar{a}\to 0$. 
While a detector at rest still has non-zero correlators $\langle \hat{Q}^2\rangle$ and $\langle \hat{P}^2\rangle$ contributed by 
vacuum fluctuations at zero temperature, such that the naive Unruh radiation $d{\cal P}^{(11)}/d\Omega_{\rm II}$ is positive as 
$\bar{a}\to 0$, the negative interference term should be able to cancel it.
Indeed, we find the radiated energy flux is largely suppressed by the interference terms when $\omega$ or $\bar{a}$ is small 
(Figure \ref{CTvarw} (left)). 
Here, for a fixed $a_0$, a smaller $\omega$ on the one hand gives a smaller averaged proper acceleration $\bar{a}$, on the other hand
it implies a longer period of oscillatory motion, so that the detector has more time to approach to the late-time state studied in 
Ref.~\cite{LH06}. Both suppress the radiated energy.


In contrast, as $\omega$ or $\bar{a}$ increases, the importance of the interference terms decreases, and the full result of the 
radiated energy flux get closer to the naive result $d{\cal P}^{(11)}/d\Omega_{\rm II}$ (Figure \ref{CTvarw} (right)).
This suggests that the Unruh-like effect experienced by the detector could be observed in the Unruh radiation in highly non-equilibrium 
conditions, with extremely short period of oscillatory motion and extremely high averaged proper acceleration.


\begin{acknowledgments}
I thank Larry Ford, Bei-Lok Hu, Pisin Chen, Daiqin Su, Tim Ralph, and Jen-Tsung Hsiang for illuminating discussions.
This work is supported by the Ministry of Science and Technology of Taiwan under Grants No. 102-2112-M-018-005-MY3 and No. 
104-2112-M-006-015, and in part by the National Center for Theoretical Sciences, Taiwan.
\end{acknowledgments}

\end{document}